# Towards Model-informed Precision Dosing with Expert-in-the-loop Machine Learning

Yihuang Kang
*Department of Information Management*
*National Sun Yat-sen University*
Kaohsiung, Taiwan
ykang@mis.nsysu.edu.tw

Yi-Wen Chiu
*Division of Nephrology,*
*Department of Internal Medicine,*
*Kaohsiung Medical University Hospital*
Kaohsiung, Taiwan
chiuyiwen@gmail.com

Ming-Yen Lin
*Division of Nephrology,*
*Department of Internal Medicine,*
*Kaohsiung Medical University Hospital*
Kaohsiung, Taiwan
mingyenlin3@gmail.com

Fang-yi Su
*Department of Information Management*
*National Sun Yat-sen University*
Kaohsiung, Taiwan
fangyisu0106@gmail.com

Sheng-Tai Huang
*Department of Information Management*
*National Sun Yat-sen University*
Kaohsiung, Taiwan
shengtai.huang28@gmail.com

*Abstract*—Machine Learning (ML) and its applications have been transforming our lives but it is also creating issues related to the development of fair, accountable, transparent, and ethical Artificial Intelligence. As the ML models are not fully comprehensible yet, it is obvious that we still need humans to be part of algorithmic decision-making processes. In this paper, we consider a ML framework that may accelerate model learning and improve its interpretability by incorporating human experts into the model learning loop. We propose a novel human-in-the-loop ML framework aimed at dealing with learning problems that the cost of data annotation is high and the lack of appropriate data to model the association between the target tasks and the input features. With an application to precision dosing, our experimental results show that the approach can learn interpretable rules from data and may potentially lower experts' workload by replacing data annotation with rule representation editing. The approach may also help remove algorithmic bias by introducing experts' feedback into the iterative model learning process.

## I. INTRODUCTION

The accessible computation and exponential growth of data have facilitated the advances of Machine Learning (ML) applications in recent years. Deep Learning [1], a powerful marriage of Artificial Neural Networks and Representation Learning [2], has brought about breakthroughs in many real-world applications. However, Deep Learning and many ML algorithms are considered black-box models, which are difficult to provide how they arrive at a decision/prediction without further interpretations. This problem has raised people's concerns on trust, safety, nondiscrimination, and other ethical issues of ML applications. Researchers and practitioners have been working on ML interpretability and developing transparent intelligent applications [3], [4]. Today, unfortunately, machines are still not as intelligent as we expected, and ML models may act in unpredictable ways instead of just crash when given unexpected inputs or asked to inference out-of-distribution cases/observations. Shortcut learning problems [5], which are decision rules that perform well on standard benchmarks but fail when applied to more challenging testing conditions, have recently drawn researchers' attention, as ML systems have been found sharing such common problems with human learning systems. Take a Melanoma (skin cancer) image classification model with deep convolutional neural networks (CNN) [6] as an example, if most of the malignant tumor images used to train the CNN models contain rulers or size markers as shown in Fig. 1, the learning algorithm may learn a shortcut that "ruler denotes skin cancer" [6].

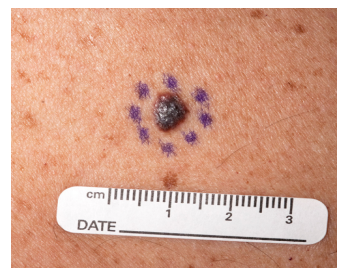

Fig. 1: A common image data used in learning Melanoma classifier (Source: National Cancer Institute)

The ML models with learned shortcuts often result in high accuracy of prediction and make questionable tasks appear solvable. Such problems may not matter when the ML system is recommending inappropriate movies to watch, but the stakes are much higher when the system is driving our car or deciding whether a patient in a hospital needs an immediate blood transfusion. Again, the ML models may perform badly on extrapolation problems or perfectly with learned shortcuts. It is clear that the machines/models are not as smart as we expect. If we do not design ML systems for human interpretation and interaction, we may find ourselves with less and less control over such data applications that govern our lives.

We here argue that humans must be a part of the ML process. In this paper, we discuss ML processes that incorporate humans into the model learning loops—Human-In-The-Loop (HITL) ML [7]. It is commonly believed that the hardest part of building ML applications is to deal with learning from limited information or labels. One simple solution is to recruit people to annotate/label datasets, but it is tedious and time-consuming to label a large amount of data. The HITL ML introduces a framework that only generates unconfident/uncertain data (edge cases) to label, which may facilitate the data annotation process. However, when building a ML application aimed at solving tasks that usually require domain experts to make complex decisions, it is difficult and expensive to ask a group of professionals to annotate thousands of data points. Some researchers believe that the HITL ML process should generate better data representations (or models) for humans to review, and such process should directly optimize for human interpretations instead of commonly-recognized error/loss measurements [8]–[11]. In this paper, we also support the arguments and propose an expert-in-the-loop ML framework in which experts can suggest how to improve the prediction, interpretation, and data representation, by giving feedback in the form of directly editing learned rules (new data representations) generated by the models and/or annotating observations/predicted values as typical HITL ML does in the model learning iterations.

We demonstrate our proposed approach with an application to model-informed precision dosing (MIPD), which refers to the modeling with the goal to help better improve a patient's health outcome by selecting personalized drug dosage regimen [12], [13]. MIPD is considered a difficult task as we usually have partial information about a patient and characteristics of patients vary significantly. The models with learned shortcuts even aggravate the non-transparency problem of the black-box models. For a successful MIPD application, we are in need of models able to withstand the scrutiny of domain experts (medical specialists). The proposed expert-in-the-loop approach is our attempt to solve the problem, by:

1. lowering user/expert workload, correcting algorithmic bias, and avoiding the shortcut learning problems;
2. modeling when the association between the target tasks and the input features are not clear, or when we have insufficient/limited data to model the associations;
3. modeling in a way that human knowledge and side information can be further encoded in successive models by evaluating the rules learned from models as well as feedback from the human experts.

The rest of this paper is organized as follows. We review background related to representation learning, ML interpretability, and the HITL ML in Section 2. The proposed approach, rule representation learning with expert-in-the-loop, is illustrated in Section 3. In Section 4, we present our experimental results with an application to precision dosing, and show our attempts to find models that optimize for human interpretation. We conclude and summarize our findings in Section 5.

## II. BACKGROUND AND RELATED WORK

The recent cheap computation, exponential growth of data, and the introduction of deep model architecture [14] have brought about breakthroughs in the development of AI and ML applications. ML algorithms have been changing almost every corner of our world. From complicated personalized medications to simple trip planning, these algorithms are playing key roles in our daily lives. Deep Learning [15] and Representation Learning [2] introduce a concept of model architecture design that promotes learning data representations that better describe the associations between the input features and the output tasks in an automatic and layer-wise fashion. Yet, the multi-modal [16] and multi-tasking [17] learning have further enhanced the architecture design by incorporating a model learning paradigm that leverages multimodal information that may potentially improve the generalization performance of multiple tasks. These ML applications are very successful in solving many machine perceptual tasks, such as visual perception [18] and speech recognition [19]. However, these complicated ensemble models have long been criticized for being black-box [3], [4], which refers to the kinds of models unable to provide explanations about how they arrive at predicted results without further interpretations. This problem has drawn attention of ML communities, and researchers/practitioners have been working on ML interpretability and developing Explainable AI [20], [21], which aims at removing biases and promoting fair, accountable, transparent, and ethical intelligent applications. Instead of focusing solely on the easy-to-quantify performance measures, such as the accuracy of prediction and the number of parameters, some researchers argue that ML models should be optimized for interpretability and humans should be a part of the model learning loop [10], [11].

In this paper, we propose a ML framework that supports the idea of the Human-In-The-Loop [7] and algorithm-in-the-loop [22] ML. However, we consider a case that we already have some labeled data and the cost of data annotation is high. Today, many supervised learning algorithms with deep architecture [23], such as deep neural networks [13] and deep forest [24], are data hungry and often require a large amount of labeled data to achieve better model performance [25]. Unfortunately, it is usually difficult to recruit appropriate data annotators and also costly to obtain high-quality labeled data. To solve the problem, researchers have proposed to incorporate humans into the model learning process—a practice of the HITL ML [7], which considers how humans and ML systems interact to solve human-machine learning problems, with a goal to shorten the time and reduce the cost to obtain deployable models [26]. In Fig 2, we show a simple HITL supervised learning process that predicts data labels.

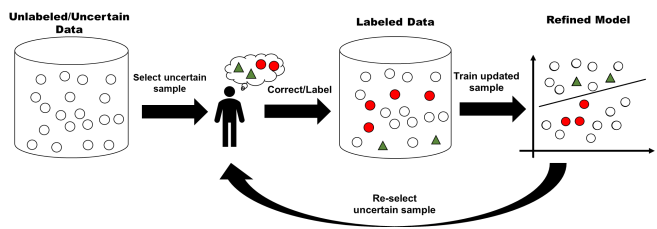

Fig. 2: A human-in-the-loop supervised learning process for classification

In many real-world ML tasks, we often have datasets with inaccurate or limited number of labels, and therefore the models learned from the data do not perform well as expected. The HITL ML provides a solution to create a model learning loop in which machines select data points with uncertain/unconfident labels and then humans annotate the selected data for the learning algorithm so as to train new models in the successive model fitting and data re-selection iterations. In Fig. 3, we illustrate common types of annotations for image data as an example adapted from [7].

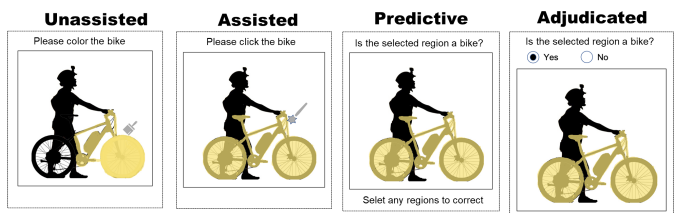

Fig. 3: Different interface designs for image data annotation (adapted from [7] with permission)

In the unassisted and assisted annotations, the annotators can manually select the data/objects with and without the help of the system (e.g., smart selection tools). The annotators have some control over the selection, but it is time-consuming for annotators when data representations are confusing (e.g., visual objects are unclear). On the other hand, in the predictive and adjudicated annotations, the system generates predictions of the data instance (e.g., object regions) and the annotators are prompted to answer a quick question with limited editing/selection options. Such annotations are least interesting tasks for annotators and often biased towards whether the annotators trust the system. Again, typical HITL ML approaches work well when the data annotation task does not involve complex decision making processes. However, when building a ML system intended to complement or replace human judgement, an interface with the aforementioned data annotation design limits the knowledge transfer from humans to machines (and vice versa) as well as creates a human-machine interactive paradigm that simply uses humans as backup. As a result, researchers argue that better data/model representations (e.g., discriminable visual objects and interpretable decision rules) can elicit better human decisions, and we should move from "learn to predict" to "learn to represent" [8], [10], [11], which consider that ML systems should be optimized to generate human-comprehensible, intermediate data representations instead of easy-to-quantify performance measurement (e.g., accuracy of predictions). We here support the argument that ML systems should be designed to reframe problems (generate better data representations) for human decision makers. But, we also consider a different real-world ML application that we train models to learn complex tasks that usually require human experts—an expert-in-the-loop ML framework.

In many domains, such as medical decision making, it is often costly and impractical to recruit a group of professionals (e.g., physicians and pharmacists) to annotate thousands of records for machines. The aforementioned data annotation strategies are not designed for efficient prior knowledge transfer from the experts. We thus propose an interactive ML framework that promotes "model annotation" (as opposed to data annotation) by directly prompting uncertain labels or new data representations in rules learned from the data for humans to annotate or edit, which aims at lowering human expert workload. Here, we discuss applications of precision dosing, which are considered complicated tasks today because they usually involve complex and important medical decisions. And the drug response is often highly variable among patients who take the same dose over time [12]. We discuss details about such applications and illustrate our proposed framework in the following sections.

## III. RULE REPRESENTATION LEARNING WITH EXPERT-IN-THE-LOOP

Consider a supervised learning task that learns a model $f$ from data consisting of $n$ observations by $p$ input features $(x_1, x_2, ... x_p) \in X$ and $m$ outputs/targets $(y_1, y_2, ... y_m) \in Y$. The model does not perform well as expected in terms of model accuracy and interpretation due to the lack of labeled $Y$ or limited information in the predictive features $X$ about the outcomes/targets $Y$. Also, the representational objects of the model (e.g., plots, statistics, and rules) used to describe the associations between $X$ and $Y$ are not optimized for humans to easily comprehend yet. To avoid asking human experts to annotate too many data instances and limiting the knowledge transfer because of improper data representations (as typical HITL ML does), we propose to create a model learning loop that iteratively re-trains models able to learn IF-THEN rule as the new data representations for humans, as well as help from humans to fine-tune the model parameters, representations, and interpretations.

As shown in Fig. 4, we first train models with the raw data to generate a set of rules $(r_1, r_2, ... r_i) \in R$ in propositional logic along with rule-encoded data points (i.e., a data point can be simply represented by the rule $r_i$ with conditions it meets) for human experts. The experts may directly edit any rules and the system could generate/sample encoded data with the corresponding predicted values $\hat{y}_{n,m}$ for experts to review. Note that, as experts may respond differently to the same rule $r_i$ and the predicted value $\hat{y}_{n,m}$, a series of inter-rater and/or intra-rater agreement tests should be conducted [26] to ensure the reliability of experts' annotations/feedback. Therefore, here we would like experts to provide their advice $a_i$ on how to modify the rules $r_i$ and the predicted values generated by the system, instead of manually to provide experts' own rules/opinions. Next, the human experts' judgement $h$ that generates advice $a_i$ (annotated rules and data points) are recorded and used as new feature extractors and new synthetic data that generate new human-centric, interpretable representations that may potentially improve the models in the following model re-training iterations. Also notice that the experts may suggest adding new features (or feature extractors) associated with the target $Y$. Again, the goal is to learn human-comprehensible and machine-processable representations while maintaining or lowering the model generalization error/loss, as defined:

$$min \sum_{i=1}^{n} loss(y_{i,m}, a_i) \text{ , for } a_i = h(x_i, \hat{y}_{i,m}, r_i)$$

where $h$ is the human judgement that gives an advice $a_i$ on a tuple consisting of the input feature vector $x_i$, the predicted value $\hat{y}_{n,m}$, and the rules $r_i$ applied to the observation $i$ for the target/output $m$. The loop continues until we obtain a model that optimizes for human interpretation—a deployable model that can be fully understood by humans. As discussed, the proposed framework is designed to help build models able to incorporate knowledge provided by human experts. Instead of simply labeling data points with a model training loop, we propose to annotate representations/rules so as to lower human workload and let machines do the tedious new representation learning and generation tasks. In the next section, we demonstrate that our framework is practicable by showing experimental results with the application to precision dosing.

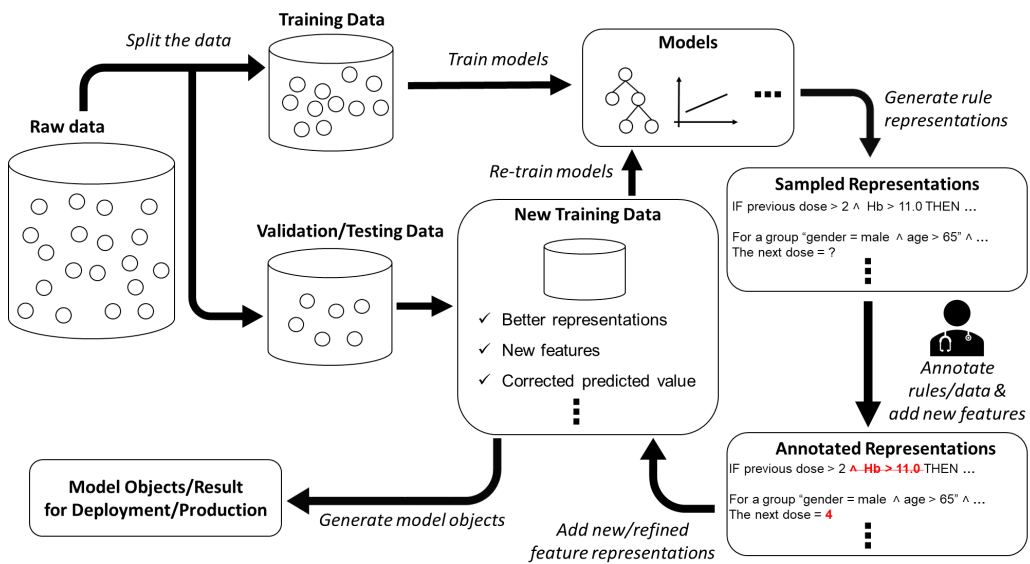

Fig. 4: The proposed representation learning with expert-in-the-loop

## IV. EXPERIMENTS AND DISCUSSION

To demonstrate our proposed approach, we used a longitudinal patient-level dataset (EPO_data) from December 2013 to May 2020 provided by the division of Nephrology in Kaohsiung Medical University Hospital (KMUH). There are a total 283 patients and 25,979 observations. The goal of using EPO_data is to show whether our proposed approach can help experts (doctors) understand more about the algorithmic decision-making process that describe the dose-response relationship as well as to improve the model interpretation while maintaining or lowering model errors with experts in the model learning loop. EPO_data contains patients' information for each visit, including patient identification (ID), timestamp, demographic information, and lab data. Fig. 5 shows an excerpt of the dataset.

| | ID | Care_Date | Hb | ... | EPO_dose | Previous_EPO_dose |
|---|---|---|---|---|---|---|
| 1 | 0001 | 2013-12-20 | 9.5 | ... | 4 | 0 |
| 2 | 0001 | 2014-01-03 | 10.8 | ... | 4 | 4 |
| 3 | 0001 | 2014-01-24 | 12.2 | ... | 1 | 4 |
| ... | ... | ... | ... | ... | ... | ... |
| n | 5211 | 2020-05-28 | 11.7 | ... | 1 | 3 |

(n = 25,979 observations; 121 features)

Fig 5: An excerpt of EPO_data

Here, we would like to build models that describe the associations between the use of erythropoietin (EPO) supplements and the changes of hemoglobin values (ΔHb) for patients in End-Stage Renal Disease (ESRD). Anemia is a common problem of the ESRD patients, and insufficient production of EPO by the kidneys is considered to be one of its major causes. One approach to treat anemia is the use of EPO-stimulating agents. Hb in dialysis patients have to be kept within the level between 10.5-11.5 mg/dl for better prognosis. The EPO supplement is a key to help keep patients' Hb in a normal range (10.5 to 11.5 is considered normal for most dialysis patients). We consider the changes of Hb values (ΔHb) as the outcome/target variable. Note that we here are dealing with correlated data with multi-dimensional measurements and correlated target variables. A mixed-effects design of the model architecture may result in better model performance in terms of model prediction and interpretation.

We first split the data into training ($n = 23,311$) and testing sets ($n = 2,668$) as shown in Fig. 6.

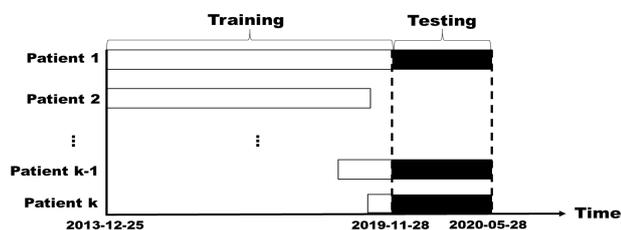

Fig 6: Train-Test split for EPO_Data

Note that we empirically partitioned the data by a certain date so that we can evaluate such mixed-effects models when they are used to recommend/inference appropriate EPO doses for patients with different lengths of medical history records. Different train-test splitting strategies with multiple testing periods/data, such as sliding windows settings, may also be used to produce more robust estimates of the generalization errors. We then fit Linear Mixed-effects Models with the best feature subset selection [28], Classification and Regression Trees (CART), Random Forest, Generalized Linear Mixed-effects Tree (GLMM Tree) [29], and bootstrap aggregating (bagging) [30] GLMM Tree. A random-intercepts Long Short-Term Memory (LSTM) neural network [1] is also designed for the performance comparison purpose. As shown in Fig. 7, the network has two inputs—the patient-level data as the general input features and a standalone patient ID input, which enables the network to predict the target with different intercepts that vary across different patients.

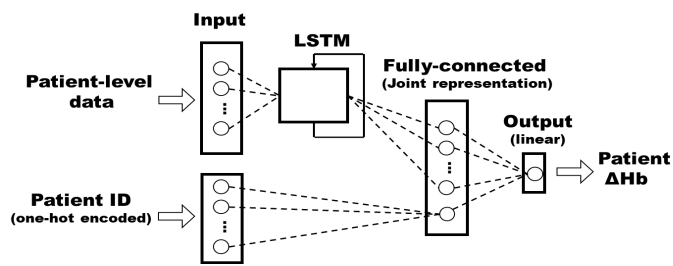

Fig. 7: LSTM neural network architecture with random intercept design

All the experiments were performed on a computing server with two Intel Xeon CPUs and a NVIDIA Geforce RTX 2080 Ti GPU, and were implemented in R 3.6.1 [31] with R package keras [32], lme4 [33], rpart [34], ranger [35], and glmertree [36]. Table 1 shows the model performance comparison in terms of training and testing Mean Absolute Error (MAE) and Root-Mean-Square Error (RMSE). We can see that the single regression tree model (CART) performs poorly compared to those models with mixed-effects design. The bagging GLMM tree with 100 tree models outperforms other models. Also, the LSTM networks do not perform well as expected. It is most likely because the network architecture in Fig. 7 only considers the random intercepts. A better network architecture that takes both fixed and random effects may further improve the performance of the networks.

Table 1: Model performance in MAE/RMSE

| Model | Training MAE | Testing MAE | Training RMSE | Testing RMSE |
|---|---|---|---|---|
| Classification and Regression Tree | 0.531 | 0.563 | 0.699 | 0.739 |
| Random Forest (# of trees = 500) | 0.483 | 0.513 | 0.640 | 0.670 |
| Linear Mixed-Effects Model | 0.465 | 0.532 | 0.613 | 0.698 |
| Linear Mixed-Effects Model Tree (GLMM Tree) | 0.453 | 0.528 | 0.597 | 0.684 |
| Bagging GLMM Tree (# of trees = 100) | **0.348** | **0.496** | **0.495** | **0.649** |
| LSTM Networks (with random intercept) | 0.498 | 0.525 | 0.655 | 0.685 |

Besides, without further model explanations, we are unable to know whether these complex models, such as bagging GLMM tree and the LSTM networks, have learned the shortcuts and thus achieved higher accuracy of predictions. As discussed, we need a model that optimizes for better human understanding instead of accuracy of prediction. And we would also like both the experts and the models learn from each other in the loop. An interpretable model, such as GLMM tree, able to generate comprehensible data representations and model interpretation is required in our proposed expert-in-the-loop ML framework. Table 1 also suggests that we might be able to identify a single GLMM tree model that balances the accuracy and interpretability by further incorporating experts' feedback. The GLMM tree consists of a set of rules that split the data space (patients) into different regions (groups) in which the predictions and dose-response associations (i.e., EPO doses versus the change of hemoglobin values) can be easily comprehended. Take the first rule/group identified by the GLMM tree as an example, the rule is defined:

> RULE #1:
> IF EPO_DOSE_per_week_3_visit_before $\leq$ 0 $\wedge$
>     PRE_Hbc_to_11_2_visit_before $\leq$ 0.3 $\wedge$
>     SUCROFER_DOSE_prev_visit $\leq$ 0
> THEN $\Delta \hat{Hb}$=−0.3306171 + 0.2261024 ∗ $EPO\_DOSE$

where the association between the EPO dose and the change of hemoglobin value (ΔHb) for the group of patients who meet the rule can be represented by a simple linear model. It is certain that the rule may be further fine-tuned by experts for better human interpretation. The experts/nephrologists have confirmed that the rules learned from the model, such as whether the patient has received Sucrofer injection, meet their expectations. However, they are also concerned that a few of the rules are questionable and the accuracy of prediction of the single GLMM tree model is slightly lower than what nephrologists could do. It is most likely because the model merely learned from the data we have without any other side information about the patients. For example, the models are unable to learn whether a patient has internal bleeding simply because we do not have the information in the data. Therefore, we believe that the proposed experts-in-the-loop framework may solve the problems and further improve the model accuracy and interpretation.

Next, we would like experts' advice on the learned new data representations (rules and predicted values) for the successive model learning iterations. As discussed previously, before further learning loop, reliability tests should be performed to ensure that there is a consensus on the patients' EPO dose-response among the nephrologists. We created an interactive web application for experts/nephrologists to annotate data and rules. As shown in Fig. 8, the nephrologists can see the groups of interest with the conditions the patients met on the left to select a particular patient and then check the predicted Hb values given different EPO doses. We conducted an experiment with five nephrologists to test the inter-rater agreement as well as to record their feedback (suggested EPO dose value) for the patients. The experts were asked to provide their own dose values for a given set of patients (n = 280 to meet the power of the reliability test). The result shows that the nephrologists do have a substantial agreement (Krippendorff's α = 0.8037, 95% CI: 0.7758-0.8279) on the prescription of the EPO, which indicates that the experts' labeled data (advice) may contain certain rules (knowledge) able to help improve the model. With the goal to further refine the learned rule representations, we then re-trained the GLMM tree with the new records and found that some rules are similar but are closer to the nephrologists' decision making processes. Take two similar rules, #28 and #21, from the previous and the new GLMM tree model respectively as an example. We can see from below that the model has identified a new condition "SUCROFER_DOSE_prev_visit" as well as updated the model parameters.

> RULE #28:
> IF EPO_DOSE_per_week_3_visit_before > 0.125 $\wedge$
>     ΔEPO_DOSE_2_visit_before > 0 $\wedge$
>     ΔHb_1_visit_before $\leq$ 1.6 $\wedge$
>     ΔHb_2_visit_before > −0.1 $\wedge$
> THEN $\Delta \hat{Hb}$=−0.4572429 + 0.2532219 ∗ $EPO\_DOSE$

> RULE #21 (new model):
> IF EPO_DOSE_per_week_3_visit_before > 0.2 $\wedge$
>     ΔEPO_DOSE_2_visit_before > 0 $\wedge$
>     ΔHb_1_visit_before $\leq$ 1.6 $\wedge$
>     SUCROFER_DOSE_prev_visit $\leq$ 0
> THEN $\Delta \hat{Hb}$=−0.6056191 + 0.2510769 ∗ $EPO\_DOSE$

Whether a patient has received Sucrofer injection in a previous visit does play a key role in the increase of his/her hemoglobin value. Such loops of rule representation annotation by experts and model re-training by machines may continue until the new representations are considered justified, which therefore could result in a deployable model that all the algorithmic decision-making processes can be comprehended by humans. We would seek to continue the loops with the help of the experts and conduct more user experiments about the interface design for rule representation annotation in our future research.

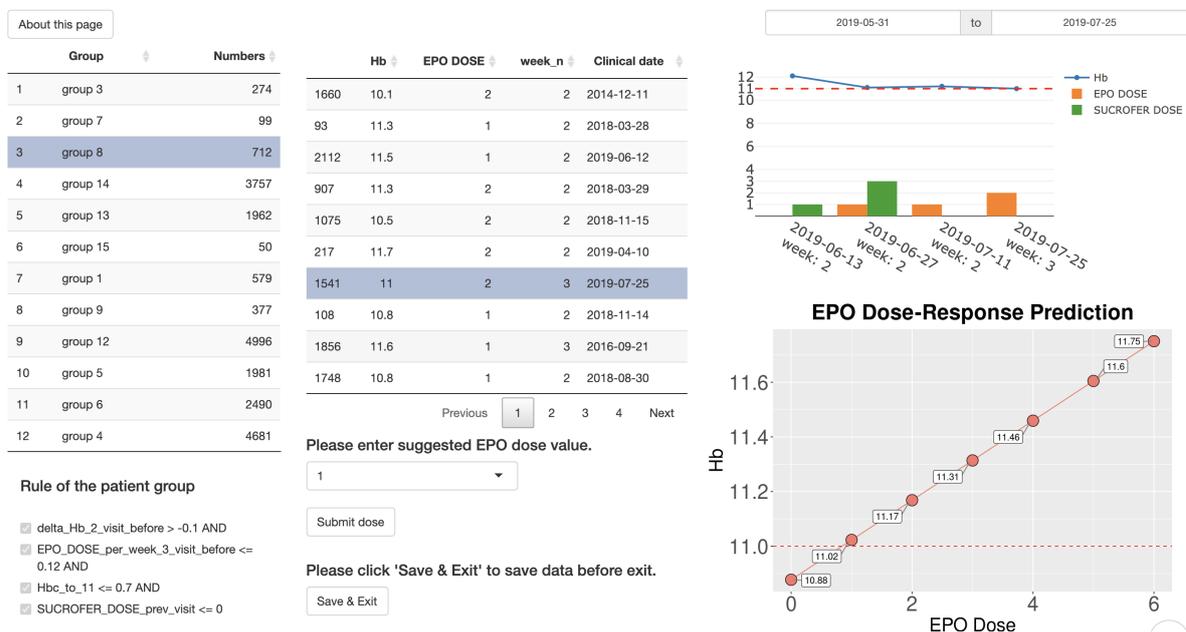

Fig. 8: Annotation interface for experts

## V. Conclusion

We propose a novel expert-in-the-loop ML framework based on an iterative and interactive rule learning/annotation process aimed at solving the problems that the cost of the data annotation is high and the lack of labeled data to model the association between the target tasks and the input features. Our experimental result on a real-world precision dosing application shows that the approach may lower the experts' workload and avoid algorithmic bias as well as undefined shortcuts by incorporating annotated rule representations into a model re-training loop.

## References

[1] I. Goodfellow, Y. Bengio, A. Courville, and Y. Bengio, Deep learning, vol. 1. MIT press Cambridge, 2016.

[2] Y. Bengio, A. Courville, and P. Vincent, "Representation learning: A review and new perspectives," IEEE Trans. Pattern Anal. Mach. Intell., vol. 35, no. 8, pp. 1798–1828, 2013.

[3] R. Guidotti, A. Monreale, S. Ruggieri, F. Turini, F. Giannotti, and D. Pedreschi, "A Survey of Methods for Explaining Black Box Models," ACM Comput Surv, vol. 51, no. 5, p. 93:1-93:42, Aug. 2018, doi: 10.1145/3236009.

[4] A. Adadi and M. Berrada, "Peeking Inside the Black-Box: A Survey on Explainable Artificial Intelligence (XAI)," IEEE Access, vol. 6, pp. 52138–52160, 2018, doi: 10.1109/ACCESS.2018.2870052.

[5] R. Geirhos et al., "Shortcut learning in deep neural networks," Nat. Mach. Intell., vol. 2, no. 11, Art. no. 11, Nov. 2020, doi: 10.1038/s42256-020-00257-z.

[6] C. Curiel-Lewandrowski et al., "Artificial Intelligence Approach in Melanoma," in Melanoma, D. E. Fisher and B. C. Bastian, Eds. New York, NY: Springer, 2019, pp. 599–628. doi: 10.1007/978-1-4614-7147-9_43.

[7] R. (Munro) Monarch, Human-in-the-Loop Machine Learning: Active learning and annotation for human-centered AI. Manning Publications, 2021.

[8] B. Wilder, E. Horvitz, and E. Kamar, "Learning to Complement Humans," Jul. 2020, vol. 2, pp. 1526–1533. doi: 10.24963/ijcai.2020/212.

[9] J. J. Dudley and P. O. Kristensson, "A Review of User Interface Design for Interactive Machine Learning," ACM Trans. Interact. Intell. Syst., vol. 8, no. 2, p. 8:1-8:37, Jun. 2018, doi: 10.1145/3185517.

[10] S. Hilgard, N. Rosenfeld, M. R. Banaji, J. Cao, and D. C. Parkes, "Learning Representations by Humans, for Humans," ArXiv190512686 Cs Stat, Oct. 2020, Accessed: Nov. 18, 2020. [Online]. Available: http://arxiv.org/abs/1905.12686

[11] I. Lage, A. Ross, S. J. Gershman, B. Kim, and F. Doshi-Velez, "Human-in-the-loop interpretability prior," in Advances in neural information processing systems, 2018, pp. 10159–10168.

[12] T. M. Polasek, S. Shakib, and A. Rostami-Hodjegan, "Precision dosing in clinical medicine: present and future," Expert Rev. Clin. Pharmacol., vol. 11, no. 8, pp. 743–746, Aug. 2018, doi: 10.1080/17512433.2018.1501271.

[13] A. S. Darwich et al., "Model-Informed Precision Dosing: Background, Requirements, Validation, Implementation, and Forward Trajectory of Individualizing Drug Therapy," Annu. Rev. Pharmacol. Toxicol., vol. 61, pp. 225–245, Jan. 2021, doi: 10.1146/annurev-pharmtox-033020-113257.

[14] Y. Bengio, "Learning deep architectures for AI," Found. Trends® Mach. Learn., vol. 2, no. 1, pp. 1–127, 2009.

[15] Y. LeCun, Y. Bengio, and G. Hinton, "Deep learning," nature, vol. 521, no. 7553, p. 436, 2015.

[16] T. Baltrušaitis, C. Ahuja, and L.-P. Morency, "Multimodal machine learning: A survey and taxonomy," IEEE Trans. Pattern Anal. Mach. Intell., vol. 41, no. 2, pp. 423–443, 2018.

[17] Y. Zhang and Q. Yang, "A Survey on Multi-Task Learning," ArXiv170708114 Cs, Jul. 2018, Accessed: Dec. 21, 2020. [Online]. Available: http://arxiv.org/abs/1707.08114

[18] K. Simonyan and A. Zisserman, "Very deep convolutional networks for large-scale image recognition," ArXiv Prepr. ArXiv14091556, 2014.

[19] A. Hannun et al., "Deep speech: Scaling up end-to-end speech recognition," ArXiv Prepr. ArXiv14125567, 2014.

[20] F. Doshi-Velez and B. Kim, "Towards A Rigorous Science of Interpretable Machine Learning," Feb. 2017, Accessed: Dec. 10, 2018. [Online]. Available: https://arxiv.org/abs/1702.08608

[21] D. Gunning, "Explainable artificial intelligence (xai)," Def. Adv. Res. Proj. Agency DARPA Nd Web, 2017.

[22] B. Green and Y. Chen, "The Principles and Limits of Algorithm-in-the-Loop Decision Making," Proc. ACM Hum.-Comput. Interact., vol. 3, no. CSCW, p. 50:1-50:24, Nov. 2019, doi: 10.1145/3359152.

[23] Y. Bengio and O. Delalleau, "On the expressive power of deep architectures," in International Conference on Algorithmic Learning Theory, 2011, pp. 18–36.

[24] Z.-H. Zhou and J. Feng, "Deep forest: towards an alternative to deep neural networks," in Proceedings of the 26th International Joint Conference on Artificial Intelligence, Melbourne, Australia, Aug. 2017, pp. 3553–3559.

[25] G. Marcus, "Deep Learning: A Critical Appraisal," ArXiv180100631 Cs Stat, Jan. 2018, Accessed: Dec. 23, 2020. [Online]. Available: http://arxiv.org/abs/1801.00631

[26] D. Xin, L. Ma, J. Liu, S. Macke, S. Song, and A. Parameswaran, "Accelerating Human-in-the-loop Machine Learning: Challenges and Opportunities," in Proceedings of the Second Workshop on Data Management for End-To-End Machine Learning, New York, NY, USA, Jun. 2018, pp. 1–4. doi: 10.1145/3209889.3209897.

[27] Y. Zhang, E. Coutinho, B. Schuller, Z. Zhang, and M. Adam, "On rater reliability and agreement based dynamic active learning," in 2015 International Conference on Affective Computing and Intelligent Interaction (ACII), 2015, pp. 70–76.

[28] D. Bertsimas, A. King, and R. Mazumder, "Best Subset Selection via a Modern Optimization Lens," ArXiv150703133 Math Stat, Jul. 2015, Accessed: Dec. 07, 2020. [Online]. Available: http://arxiv.org/abs/1507.03133

[29] M. Fokkema, N. Smits, A. Zeileis, T. Hothorn, and H. Kelderman, "Detecting treatment-subgroup interactions in clustered data with generalized linear mixed-effects model trees," Behav. Res. Methods, vol. 50, no. 5, pp. 2016–2034, Oct. 2018, doi: 10.3758/s13428-017-0971-x.

[30] L. Breiman, "Bagging Predictors," Mach. Learn., vol. 24, no. 2, pp. 123–140, Aug. 1996, doi: 10.1023/A:1018054314350.

[31] R Core Team, R: A Language and Environment for Statistical Computing. R Foundation for Statistical Computing, 2019. [Online]. Available: https://www.R-project.org/

[32] F. Chollet and J. J. Allaire, R interface to keras. GitHub, 2017. [Online]. Available: https://github.com/rstudio/keras

[33] D. Bates et al., lme4: Linear Mixed-Effects Models using "Eigen" and S4. 2020. Accessed: Dec. 10, 2020. [Online]. Available: https://CRAN.R-project.org/package=lme4

[34] T. M. Therneau, E. J. Atkinson, and others, An introduction to recursive partitioning using the RPART routines. Technical Report 61. URL http://www. mayo. edu/hsr/techrpt/61. pdf, 1997. Accessed: May 12, 2016. [Online]. Available: http://r.789695.n4.nabble.com/attachment/3209029/0/zed.pdf

[35] M. N. Wright and A. Ziegler, "ranger: A fast implementation of random forests for high dimensional data in C++ and R," ArXiv Prepr. ArXiv150804409, 2015.

[36] M. Fokkema and A. Zeileis, glmertree: Generalized Linear Mixed Model Trees. 2019. Accessed: Dec. 10, 2020. [Online]. Available: https://CRAN.R-project.org/package=glmertree